\documentclass[a4paper,11pt]{article}
\pdfoutput=1 

\usepackage{jcappub} 

\usepackage{geometry} 
\usepackage{ulem}
\usepackage{enumitem}
\usepackage{graphicx}
\usepackage{booktabs}
\usepackage{amsmath}
\usepackage{amssymb}
\usepackage{todonotes}
\usepackage{color}
\newcommand{\refeq}[1]{Eq.~(\ref{eq:#1})}

\title{Bayesian Inference Constraints on Astrophysical Production of Ultra-high Energy Cosmic Rays and Cosmogenic Neutrino Flux Predictions}
\author[1]{Andr\'es Romero-Wolf}
\author[2]{M\'aximo Ave}

\affiliation[1]{Jet Propulsion Laboratory, California Institute of Technology, Pasadena, CA 91109}
\affiliation[2]{ Instituto de F\'isica de S\~ao Carlos, Universidade de S\~ao Paulo, CP 369, 13560-970, S\~ao Carlos, SP, Brasil}

\emailAdd{Andrew.Romero-Wolf@jpl.nasa.gov}

\abstract{
A flux of extra-terrestrial neutrinos at energies $\gg10^{15}$~eV has the potential to serve as a cosmological probe of the high-energy universe as well as tests of fundamental particle interactions. Cosmogenic neutrinos, produced from the interactions of ultra-high energy cosmic rays (UHECRs) with cosmic photon backgrounds, have been regarded as a guaranteed flux. However, the expected neutrino flux depends on the composition of UHECRs at the highest energies; heavier nuclei result in lower neutrino fluxes compared to lighter nuclei and protons.
The objective of this study is to estimate the range of cosmogenic neutrino spectra consistent with recent cosmic-ray spectral and compositional data using a fully inferential Bayesian approach. The study assumes a range of source distributions consistent with astrophysical sources, the flux and composition of cosmic rays, and detector systematic uncertainties. 
The technique applied to this study is the use of an affine-invariant Markov Chain Monte Carlo, which is an effective Bayesian inference tool for characterizing multi-dimensional parameter spaces and their correlations. 
}

\begin{document}
\maketitle

\tableofcontents
\pagebreak

\section{Introduction}
The possibility of discovering new astrophysical phenomena through the observation of high-energy neutrinos has inspired a number of recent experimental efforts~\cite{Gorham_2004, Scholten_2009, James_2010, Kravchenko_2006, Gorham_2010, IceCube_UHE_2016, ARA_2012, Martineau-Huynh_2017, Auger_Neutrino_2015}. Unlike high-energy photons and baryons, high-energy neutrinos can traverse the cosmos without suffering significant energy losses and thus provide a probe into high energy-phenomena throughout the entire Universe. The study of high-energy cosmic neutrinos also offers the opportunity to test particle interactions and fundamental laws~\cite{Cornet_2001, Jain_2002, Reynoso_2013}. However, the fluxes of plausible astrophysical sources are expected to be challengingly low and uncertain~\cite{Kotera_2010}. 

The existence of Ultra High Energy Cosmic Rays (UHECRs $E\gtrsim10^{19}$~eV), together with their well established interactions with photon background fields, have led astrophysicists to expect that high-energy neutrino fluxes (above EeV) should be {\it guaranteed} at some level.  These neutrinos could be produced at the source due to interactions of cosmic ray particles during acceleration ({\it source} neutrinos)~\cite{Waxman_Bahcall_1999} or during propagation of cosmic rays from the sources to Earth ({\it cosmogenic} neutrinos)~\cite{Berezinsky_Zatsepin_1969}. {\it Source} neutrino fluxes are difficult to predict since a detailed knowledge of the  astrophysical environment is required.  {\it Cosmogenic} neutrinos are often considered {\it guaranteed} since their flux level is directly related with the observed flux level of cosmic rays. The goal of this work is to assess the level of the {\it cosmogenic} neutrino flux using constraints from the recent measurements provided by UHECR experiments.

Within the last decade, the Pierre Auger Observatory (PAO) detected a cutoff in the cosmic-ray spectrum above $4 \times10^{19}$~eV~\cite{Abraham_2008} while the Telescope Array (TA) experiment has observed the same spectral feature at energies above $5.4 \times10^{19}$~eV~\cite{Abu-Zayyad_2013}.
The discrepancy in the energy of the cutoff is consistent within the systematics of each experiment~\cite{Ivanov_2017}. In addition, both PAO~\cite{Aab_2014} and TA~\cite{Belz_2015} have produced data on the depth of shower maximum $X_\mathrm{max}$ as a function of energy, which is a tracer of cosmic-ray composition. 
A recent joint study by TA and Auger has concluded that the mean value and root-mean-square of $X_\mathrm{max}$ are consistent within systematic uncertainties of each experiment~\cite{deSouza_2017}.
The Auger results~\cite{Aab_2014b} indicate that the composition shifts from proton-dominated to Helium or Carbon-Nitrogen-Oxygen dominated as the energy increases.

There are competing hypotheses for the observed suppression of the ultra-high energy cosmic ray spectrum. In 1966, Greisen~\cite{Greisen_1966} and Zatsepin \& Kuzmin~\cite{Zatsepin_Kuzmin_1966} predicted a cutoff due to interactions of cosmic-ray protons with the cosmic microwave background (CMB) radiation. Alternatively, the observed cutoff could also be due to the cosmic ray source energy available to particle acceleration~\cite{Aloisio_2011}. In this study we assume a rigidity ansatz, where the source has a mixed composition of protons and heavier nuclei with the energy cutoff of each element by $ZE_\mathrm{max}$, where $Z$ is the nuclear charge and $E_\mathrm{max}$ is the maximum acceleration energy of the proton. The observed cosmic-ray spectrum is a combination of the source acceleration energy, propagation effects of cosmic rays, and the composition of the sources. 

In a recent study, the Auger collaboration produced a combined fit of the spectrum and composition data measured by the PAO~\cite{PAO_2017}. They applied a Bayesian approach, including the systematic uncertainties of their detectors, to study the results for several propagation, source evolution, and hadronic interaction models. Despite the variety of results on the inferred composition and source spectrum, the best-fit energy cutoff for protons, or rigidity cutoff, was, in the majority of cases, found to be consistently between 1.4 -- 78~EeV, a range which contains the GZK interaction threshold ($\sim 40$~EeV) and values well below that threshold.

The expected flux of neutrinos produced by interactions of cosmic rays propagating through the cosmic photon background (infrared, CMB, and radio) were first predicted by Berezinsky and Zatsepin in 1969~\cite{Berezinsky_Zatsepin_1969}. These cosmogenic neutrinos are preferentially produced for protons above $4\times 10^{19}$~eV, which is the $\Delta^+$ resonance threshold with a CMB photon. However, they can also be produced by interactions of nuclei and other photon backgrounds at a reduced level due to photo-disintegration and nuclear decay. Several studies have produced expectations of the cosmogenic neutrino flux assuming the cosmic-ray spectrum is proton-dominated up to the highest energies observed $\gtrsim 10^{20}$~eV \cite{Engel_Seckel_Stanev_2001, Kalashev_2002, Barger_2006, Stanev_2006}, intermediate composition~\cite{Kotera_2010}, and,  in pessimistic cases, pure Iron~\cite{Ave_2005}.
There is a large range of uncertainty in the flux not only due to composition, but also due to assumptions in the cosmic photon backgrounds, interactions, source spectrum, and source evolution (see, for example,~\cite{Alves_Batista_2015} and references therein). 

Neutrino flux estimates from the studies mentioned above have motivated several searches for neutrinos with energies $>10^{16}$~eV using the Moon~\cite{Gorham_2004, Scholten_2009, James_2010}, the Antarctic Ice Cap~\cite{Kravchenko_2006, Gorham_2010, IceCube_UHE_2016} with future detectors planned~\cite{ARA_2012, Martineau-Huynh_2017}, and via air shower production~\cite{Auger_Neutrino_2015}. Recently an extra-terrestrial flux of neutrinos extending above $10^{15}$~eV has been detected with IceCube~\cite{IceCube_detection_2014}. However, the neutrino flux at energies $>10^{16}$~eV has not yet been detected, with the best limits being produced by IceCube~\cite{IceCube_UHE_2016}, Auger~\cite{Auger_Neutrino_2015}, and ANITA~\cite{Gorham_2010}, in different parts of the energy spectrum. 

In this study we estimate the expected range of neutrino fluxes assuming cosmic particle accelerators that are consistent with the observations of the PAO. It is important to note that cosmic-ray observations on Earth are dominated by local sources ($\lesssim 300$~Mpc). Neutrinos, on the other hand, once produced, propagate practically unimpeded throughout the Universe. It is possible that cosmic particle accelerators beyond our local Universe have a different behavior compared to local sources. However, without priors on how that behavior may change, it is important to estimate the neutrino flux resulting from the assumption that local UHECR sources are representative of particle accelerators out to higher redshifts. 

 We take a Bayesian approach using the \texttt{emcee} affine-invariant Markov Chain Monte Carlo (MCMC)~\cite{emcee}. We present marginalized posterior distributions on the cosmic accelerator models using data from the PAO including the predicted range of fluxes. The propagation of cosmic-ray nuclei and the associated neutrino production is modeled using the publicly available CRPropa3 software package~\cite{CRPropa3_2016}. The source code used in this study is also being made publicly available\footnote{\url{https://github.com/afromero/CRANE}}.

A study by Taylor, Ahlers, \& Hooper in 2015~\cite{Taylor_2015} applied a Markov Chain Monte Carlo method to fit the spectrum and composition of the Auger cosmic-ray spectrum assuming different sets of source distributions. They consider source distributions evolving as $(1+z)^{n}$ for fixed values of $n=$-6, -3, 0, and +3. Values with $n<+3$, are outside the  range expected for typical astrophysical sources (e.g. Gamma Ray Bursts, Active Galactic Nuclei, Star Formation Rate, and Faranoff-Riley type II galaxies). Models with  $n<0$ simply would imply that the UHECR spectrum is dominated by a few local sources. They found that the spectral index $\alpha$ is strongly correlated with the source evolution index, with $n$=-6 giving a best-fit spectral index of $\alpha=1.83\pm0.31$, which is consistent with values expected from shock acceleration ($\alpha\simeq$2), while $n$=+3 gave a best-fit $\alpha=0.64\pm0.44$, which is not consistent with these models. In their conclusions, they state that their results significantly favor negative source evolution indices since these result in a spectral index consistent with shock acceleration. It is important to note that several studies have proposed models with spectral indices significantly different from $\alpha\simeq2$. For example, models of newly born pulsars~\cite{Fang_2012} and tidal disruption events~\cite{Zhang_2017} as sources of cosmic rays give $\alpha\simeq1$. A model of UHECR of shear acceleration from black hole jets~\cite{Kimura_2018} gives $\alpha$ between 0 and 1. There have even been Gamma Ray Burst acceleration simulations~\cite{Globus_2015} that, in some cases, give $\alpha\simeq 0$.

In this work, we assume that the contribution to the observed flux is from astrophysical sources while not imposing prior constraints on the source spectral index. In that context, we provide  the full statistical range of astrophysical neutrino spectra consistent with the observations of the PAO. 

This paper is organized as follows. In section 2 we provide the details of our cosmic ray nuclei and neutrino flux model along with a characterization of the CRPropa3-derived particle yields. In Section 3 we present the likelihood function and forward modeling approach used for the Bayesian inference analysis. In Section 4 we present the results on marginalized posteriors of the model parameters, consistency checks with the PAO data, and neutrino flux predictions. In Section 5 we discuss the significance of the results and future prospects. 

\section{Flux Model}
The model of the cosmic ray and neutrino flux incident on Earth assumes a set of sources with luminosity function $L_{\mathrm{Z}_\mathrm{src},\mathrm{A}_\mathrm{src}}(E_\mathrm{src})$ dependent on the source particle nuclear charge and mass $(\mathrm{Z}_\mathrm{src}, \mathrm{A}_\mathrm{src})$ and energy $E_\mathrm{src}$. As the particles propagate they interact with cosmic background photons resulting in the production of secondary nuclei, neutrinos, and gamma rays. The yield function of a secondary particle species, indexed by $k$, per source particle for observed energy between $E_\mathrm{obs}$ and $E_\mathrm{obs}+dE_\mathrm{obs}$ is denoted by $Y_k(E_\mathrm{obs}|E_\mathrm{src}, \mathrm{Z}_\mathrm{src}, \mathrm{A}_\mathrm{src}, z)$ depends on the source particle species, energy, and redshift $z$. For a distribution of sources with comoving density $n_\mathrm{src}$, the total observed flux of particles of species $k$ is given by
\begin{equation}
\begin{split}
I_{k}(E_\mathrm{obs}) & = \sum_{(\mathrm{Z}_\mathrm{src}, \mathrm{A}_\mathrm{src})}
\int dz \ \frac{c}{H_0}\frac{n_\mathrm{src}(z)}{(1+z)\sqrt{\Omega_M(1+z)^3+\Omega_{\Lambda}}}  \\
& \ \ \ \ \ \ \ \ \ \ \ \ \ \ \   \int  
dE_\mathrm{src} \ 
{Y}_{k}(E_\mathrm{obs} | E_\mathrm{src}, \mbox{Z}_\mathrm{src}, \mbox{A}_\mathrm{src}, z) \ L_{\mathrm{Z}_\mathrm{src},\mathrm{A}_\mathrm{src}}(E_\mathrm{src})
\end{split}
\label{eq:FluxModel}
\end{equation} 
where 
$c$ is the speed of light, $H_0$ is the Hubble constant, 
$\Omega_M$ is the fraction energy density of matter and $\Omega_\Lambda$ is the fractional energy density of the cosmological constant. The flux is the rate of particles arriving per area element $dA$ per solid angle $d\Omega$ in the energy range between $E_\mathrm{obs}$ and $E_\mathrm{obs} + dE_\mathrm{obs}$. See Appendix A for a derivation. The validity of this model rests on the assumption that the comoving density of sources per Mpc$^{3}$ is large enough to be approximated by an integral and that the cosmic-ray spectrum is not dominated by one or a few nearby sources.

\subsection{Source Luminosity and Distribution Functions}
We parameterize the source luminosity function according to a power law with an exponential high-energy cutoff
\begin{equation}
L_{\mathrm{Z}_\mathrm{src},\mathrm{A}_\mathrm{src}}(E_\mathrm{src}) = L_{0,\mathrm{Z}_\mathrm{src},\mathrm{A}_\mathrm{src}} \left(\frac{E_\mathrm{src}}{E_\mathrm{ref}}\right)^{-\alpha}
\times
\begin{cases} 
1, & \mbox{if }E_\mathrm{src}<\mathrm{Z}_\mathrm{src}E_\mathrm{max}  
\\ 
\exp\left(1-\frac{E_\mathrm{src}}{\mathrm{Z}_\mathrm{src}E_\mathrm{max}}\right)
, & \mbox{if }E_\mathrm{src}\geq\mathrm{Z}_\mathrm{src}E_\mathrm{max}
\end{cases}
\label{eq:luminosity_model}
\end{equation}
Note this parameterization is of the same form used by the Auger collaboration in~\cite{PAO_2017} but not in Taylor 2015~\cite{Taylor_2015} .
The normalization $L_{0,\mathrm{Z}_\mathrm{src},\mathrm{A}_\mathrm{src}}$ corresponds to the luminosity of particle species $(\mathrm{Z}_\mathrm{src},\mathrm{A}_\mathrm{src})$ at fixed source reference energy $E_\mathrm{ref}$. 
The energy dependence corresponds to a power law with spectral index $\alpha$ along with a exponential cutoff. This functional shape describes a wide variety of cosmic-ray acceleration models. The cutoff energy $\mathrm{Z}_\mathrm{src}E_\mathrm{max}$, with $E_\mathrm{max}$ being the proton cutoff energy, is assumed to depend only on rigidity. This approach is valid as long as the interaction losses of different nuclear species at the source is similar or negligible. This rigidity ansatz is not the general case, see for example~\cite{Globus_2015}. In a future update of this work we will consider non-rigidity ansatz cases as well. 

The reference luminosity is given by $L_{0,\mathrm{Z}_\mathrm{src},\mathrm{A}_\mathrm{src}} = f_{\mathrm{Z}_\mathrm{src},\mathrm{A}_\mathrm{src}} L_0$, where $f_{\mathrm{Z}_\mathrm{src},\mathrm{A}_\mathrm{src}}$ is the fraction of the flux at the source reference energy due to a given nuclear species and $L_{0}$ is a global reference luminosity.
We consider primary protons, and the most abundant isotopes of Helium, Nitrogen, Silicon, and Iron. Each of these are meant to be representative fractions for groups of nuclei, as in~\cite{Taylor_2015} and~\cite{PAO_2017}. The proton fraction is labeled by $f_\mathrm{p}$ and according to the periodic table symbol for heavier nuclei with $f_\mathrm{p}+f_\mathrm{He}+f_\mathrm{N}+f_\mathrm{Si}+f_\mathrm{Fe}=1$

The comoving density of sources as a function of redshift $n_\mathrm{src}(z)$ 
is parameterized as a piece-wise power law
\begin{equation}
n_\mathrm{src}(z)=
 \begin{cases} 
      n_0 \ (1+z)^{n} & 0\leq z\leq z_1 \\
      n_0 \ (1+z_1)^{n} & z_1\leq z\leq z_2 \\
      n_0 \ (1+z_1)^{n}\exp((z_2 - z)/z_2) & z_2\leq z\leq z_\mathrm{max} \\
      0   & z > z_\mathrm{max}
   \end{cases}
\label{eqn:source_distrib}
\end{equation}
The parameter $n_0$ gives peak value of comoving density of sources, the index $n$ describes the evolution from redshift 0 and $z_1$. The source density plateaus region between $z_1$ and $z_2$ followed by an exponential fall-off for $z>z_2$. We place a hard cutoff in the source density for $z>z_\mathrm{max}$. This form of the source evolution roughly covers the shape of the source evolution of astrophysical source such as star formation rate, gamma ray bursts, and Faranoff-Riley type II galaxies (see~\cite{Kotera_2010}). 

\subsection{Summary of Source Parameters and Priors}
The product of the reference luminosity $L_0$ and peak source density $n_0$ is indiscernible. We therefore use a flux normalization parameter $N_0 = L_0n_0$. We provide a list of source model parameters and their prior range in Table~\ref{tab:parameters}.

\begin{table}[!htbp]
   \caption{Flux model parameter and bounds.}
   \centering
   \begin{tabular}{@{} lll @{}} 
      \toprule
      Parameter     & Description & Prior Bounds\\
      \midrule
      $N_0$         & Flux normalization & $N_0 > 0$ \\
      $f_i$         & Fractional composition for           &  $f_{i}\in[0,1]$, $\sum_i f_i =1$ \\
                    & primary $i\in\{\mathrm{p}, \mathrm{He}, \mathrm{N}, \mathrm{Si}, \mathrm{Fe}\}$   &  Free composition \\
      $\alpha$      & Source spectral index  & $\alpha\in[-10,10]$ \\
      $E_\mathrm{max}$     & Maximum energy of    & $E_\mathrm{max}>10^{17}$~eV \\
                    & acceleration for protons                                &   \\
      $n$           & Source evolution index  &   $n\in[3.4,5.0]$\\
      $z_1$         & Plateau redshift  &  $z_1\in[1,2.7]$ \\
      $z_2$         & Turn-around redshift  &  $z_2\in[2.7,\mathrm{min}(4,z_\mathrm{max})]$ \\
      $z_\mathrm{max}$     & Cutoff redshift   &  $z_\mathrm{max}\in[2.7,10]$ \\
      \bottomrule
   \end{tabular}
   \label{tab:parameters}
\end{table}

The overall normalization parameter is required to have $N_0>0$ since the flux is positive. The fractional composition $f_i$ for each element $i$ is required to have a sum of unity with each $f_i\geq 0$ but is otherwise unconstrained. 

We allow for the source spectral index to lie in the broad range $\alpha\in[-10,10]$. The results of~\cite{Taylor_2015} showed, that depending on the source evolution index assumed, $\alpha$ is consistent with values between -0.4 and +2.4. Although we are not using exactly the same source luminosity model as~\cite{PAO_2017}, they have shown that, within the systematic uncertainties, their best-fit models can give a source spectral index of -1.5, which is a rising power law. The gamma ray burst acceleration models of~\cite{Globus_2015} produce source luminosity spectra consistent with spectral indices $\sim 0$ and exponential cutoff. 

We require the maximum energy of acceleration for protons $E_\mathrm{max}>10^{17}$~eV. This value is well below the results of~\cite{PAO_2017} and~\cite{Taylor_2015} and not expected to strongly influence the posterior value.
For the source models, we set priors on the source evolution index $n\in[3.4,5.0]$ with $z_1\in[1,1.7]$,  $z_2\in[2.7,4]$,  $z_\mathrm{max}\in[2.7,10]$ and $z_1\leq z_2 \leq z_\mathrm{max}$. 
The previous work of~\cite{Taylor_2015} showed that the neutrino fluxes for mixed composition can change by a factor of 100 by changing $n=-6$ to $n=+3$, indicating a strong dependence on the source index prior. In this work, we focus strictly on source evolution parameters that cover some typical astrophysical source classes of interest. 

\subsection{Secondary Particle Yields}
We use the CRPropa3~\cite{CRPropa3_2016} open source code\footnote{\url{https://github.com/CRPropa/CRPropa3}} to model the propagation of UHECRs along with the production of secondary nuclei and neutrinos. The propagation includes the photo-pion production, photo-disintegration, and electron pair production of primary and secondary nuclei with the cosmic microwave background (CMB) and the Gilmore 2012~\cite{Gilmore_2012} model of the infrared background (IRB) along with nuclear decays and the adiabatic energy losses due to the expansion of the Universe. We propagate mono-energetic primary protons and the most common isotopes $_2^{4}$He, $_{7}^{14}$N, $_{14}^{28}$Si, and $_{26}^{56}$Fe to produce the particle yields $Y_k$ arriving at Earth. The neutrinos produced from photo-disintegration and nuclear decay are tracked and propagated.
Although in reality a larger variety of elements and isotopes would be accelerated, we take the set of elements listed above as representative of their groups. A similar approach was taken in~\cite{Taylor_2015} and~\cite{PAO_2017}.  

In this study we have neglected the effect of galactic and intergalactic magnetic fields. Galactic magnetic fields can affect the arrival distribution of UHECRs, but they will not modify the observed spectrum or composition. Intergalactic magnetic fields (IGMF) are bounded to be $\lesssim 10^{-9}$~G by observations of the CMB and $\gtrsim 10^{-17}$~G by time-delayed GeV $\gamma$-ray emission~\cite{Durrer_Neronov_2014}. It is expected that IGMFs increase the path length of charged nuclei at any given redshift potentially modifying both cosmic-ray energy loss processes and neutrino yields. Tracking magnetic field deflection increases the complexity of the one-dimensional cosmological evolution propagation models to a four-dimensional simulation. Turbulent IGMFs cause the propagation of UHECRs to lie in the diffusion regime below a given energy threshold, introducing a magnetic horizon that will suppress the flux, although the suppression is only significant for UHECR with $E\lesssim10^{17}$~eV~\cite{Alves_Batista_2014}. Effects of the IGMF may result in important modifications to the results of this study but will be deferred to a future publication. Recent results indicate that for the upper bound of the IGMF strength, significant modifications to the fitted spectral index and energy cutoff may occur~\cite{Wittkowski_2017}, although the fitted energy cutoff remains below 10$^{19}$~eV. 
 
\begin{figure}[htbp]
  \centering
   \includegraphics[clip,width=1.0\linewidth]{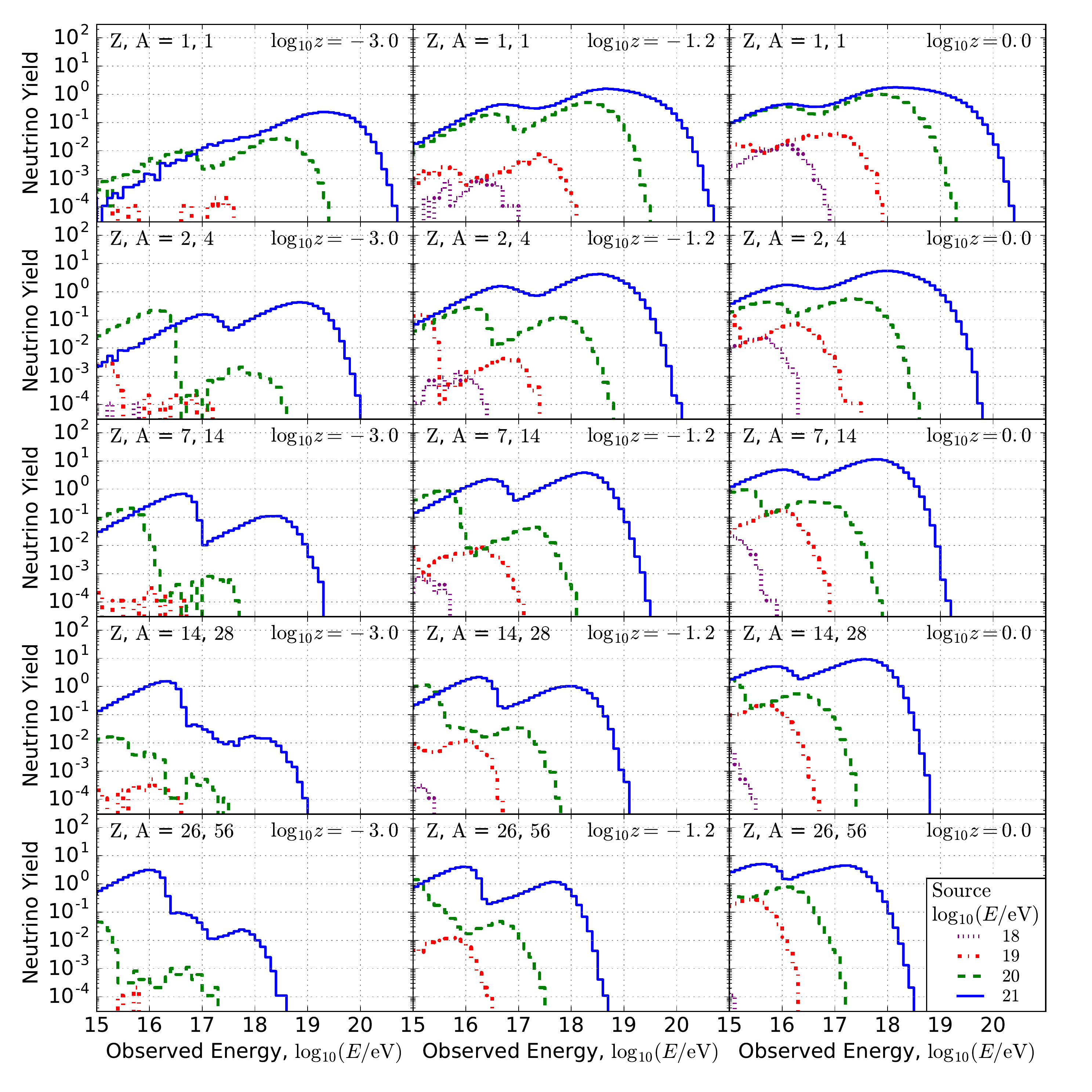} 
   \caption{Neutrino yields as a function of neutrino arrival energy for source nucleus energies of $10^{18}$, $10^{19}$, $10^{20}$, and $10^{21}$~eV. From top to bottom, the parent nuclei are protons, He, N, Si, and Fe. From left to right, the source redshift is $z$=0.001, roughly corresponding to the distance to Centaurus A (the nearest AGN), $z=0.06$, roughly corresponding to the GZK horizon, and $z$=1, where nuclear secondaries of UHECRs are not expected to arrive at Earth with ultra-high energies, but the secondary neutrinos may do so. }
   \label{fig:neutrino_yields}
\end{figure}

In Figure~\ref{fig:neutrino_yields} we show the neutrino yields for protons, He, N, Si, and Fe at redshifts corresponding to $\sim$4~Mpc (the location of Cen A, our nearest neighbor AGN), a few hundred Mpc, corresponding to the GZK horizon, and a redshift of 1, corresponding to a cosmological distance from which contribution to the observed UHECR flux is negligible but neutrino fluxes would still be produced. 
Neutrino yields increase with redshift due to the longer cosmic-ray propagation path meaning more distant objects have a higher contribution. 
The neutrino yields from different nuclear species have their cutoff shifted according to $1/\mathrm{A}_\mathrm{src}$, which suppresses the high energy flux for heavy-composition-dominated UHECR spectra.

\clearpage

\section{Likelihood Functions}

The likelihood that a set of parameters fits the data depends on comparisons between the PAO measurements of the energy spectrum and shower depth with models of the cosmic ray flux, $\langle X_\mathrm{max}\rangle$, and RMS($X_\mathrm{max}$) as a function of energy. In each case, we account for the systematic uncertainties in energy and composition using free parameters included in the likelihood function. Evaluation of the flux Equation~\ref{eq:FluxModel} with the parameters in Table~\ref{tab:parameters} provides a prediction of the observed cosmic-ray spectrum and composition. UHECR measurements for energies larger than 10$^{18.9}$~eV will be used to constrain these parameters. The reason we choose this energy is that the spectrum has a break at 10$^{18.7}$~eV that could be interpreted as the transition from a galactic to extragalactic cosmic ray flux. To avoid having to fit galactic contributions we move to a higher energy bin where a potential galactic contribution is negligible. 
In this section we describe the likelihood function used and our procedure to account for the experimental systematic uncertainties.

\subsection{Cosmic-Ray Flux}
The flux models presented above are treated as the true values given a set of parameters. From this point we forward model the cosmic-ray event counts for each energy bin observed by the PAO accounting only for a single systematic energy shift since this is the dominant term (for a more detailed treatment see~\cite{PAO_2017}). To account for systematic uncertainties in the energy, we introduce a parameter $\epsilon = \delta E / E$ that corresponds to a constant fractional energy shift between the true energy and the systematic offset in the assigned energy for the event.
Given a flux $I(E')$, the cosmic-ray counts over an energy bin of $E_{i}$ of bin width $\Delta \log_{10}E$ is given by
\begin{equation}
N_\mathrm{model}(\log_{10}E_i, \Delta \log_{10}E , \epsilon)
= 
\ln(10) T 
\int_{\log_{10}(E_i(1+\epsilon))}^{\log_{10}(E_i(1+\epsilon))+ \Delta \log_{10}E } 
d\log_{10}E' \ G(E')  \ E' I(E').
\end{equation}
where $T$ is the observation time and $G(E)$ is the \'etendu (detector area integrated over solid angle or acceptance), which is generally energy dependent. For the energies of interest ($E>10^{18.9}$~eV) 
we assume Auger has a flat response~\cite{Schulz_2013} so we can factor it out of the integral to give
\begin{equation}
N_\mathrm{model}(\log_{10}E_i, \Delta \log_{10}E , \epsilon)
= 
\ln(10) \ T G
\int_{\log_{10}(E_i(1+\epsilon))}^{\log_{10}(E_i(1+\epsilon))+ \Delta \log_{10}E } 
d\log_{10}E'   \ E' I(E').
\end{equation}

For $\epsilon = 0$, assuming a sufficiently small bin width $\Delta \log_{10}E$, the integral approximates to 
\begin{equation}
N_\mathrm{model}(\log_{10}E_i, \Delta \log_{10}E , 0)
\simeq 
\ln(10) \ T G \
\Delta\log_{10}E'   \ E_i I(E_i).
\end{equation}
When $\epsilon>0$, $N_\mathrm{model}$ is obtained by a linear combination of neighboring bins
\begin{equation}
\begin{split}
N_\mathrm{model}(\log_{10}E_i, \Delta \log_{10}E, \epsilon) & = \\
& (1-u)N_\mathrm{model}(\log_{10}E_i, \Delta \log_{10}E, 0) \\
& + u N_\mathrm{model}(\log_{10}E_i + \Delta \log_{10}E, \Delta \log_{10}E,  0) \\
\end{split}
\end{equation}
where $u=\log_{10}(1+\epsilon)$.  

Systematics in the energy scale in Auger are estimated to be $\sim14\%$~\cite{Valino_2015}. 
We model this as a Gaussian distributed random variable with $\sigma_{\epsilon}=0.14$ and we limit $\epsilon\in[-\sigma_\epsilon, +\sigma_\epsilon]$ so as not to allow the solution to significantly stray far from the experimentally constrained systematics.  

The likelihood is calculated using Poisson statistics on the bin counts
\begin{equation}
\mathcal{L} = \prod_{i} \frac{N_{i,model}^{N_{i,data}}e^{-N_{i,model}}}{N_{i,data}!}.
\end{equation}
The log likelihood is 
\begin{equation}
\ln\mathcal{L} = \sum_{i} \left[ N_{i,data}\ln(N_{i,model}) - \ln (N_{i,data}!) - N_{i,model} \right].
\end{equation}
For high counts, the likelihood can be approximated with a Gaussian distribution with mean $N_\mathrm{model}$ and standard deviation $\sqrt{N_\mathrm{model}}$.

\subsection{Cosmic-Ray Composition}
Due to the low fluxes at ultra-high energies, the detection of UHECRs can only be achieved by measuring extensive air showers (EAS), cascades of secondary particles resulting from the interaction of the primary cosmic rays with the Earth atmosphere. The position of the shower maximum ($X_\mathrm{max}$) and the muon content ($N_\mu$) are sensitive to the atomic mass of the primary that initiated the shower. Current experiments measure the position of the shower maximum with an accuracy below the expected shower-to-shower fluctuations. Therefore, the mean and root-mean-square (RMS) of $X_\mathrm{max}$ are sensitive to the primary composition. Another possible approach is to fit the composition using likelihood functions on simulated $X_\mathrm{max}$ distributions as was done in~\cite{Aab_2014}. However, for simplicity, we have chosen to model the mean and RMS of the $X_\mathrm{max}$ distributions.

We use the CONEX Monte Carlo~\cite{Bergmann_2007} to compute the energy dependence of the mean and RMS of $X_\mathrm{max}$ as a function of energy and primary mass. The QGSJetII-04~\cite{Ostapchenko_2011} and EPOS-LHC~\cite{Pierog_2015} hadronic model generators are used. These models predict the same energy dependence of the mean and RMS of the  $X_\mathrm{max}$ distribution, but with an overall systematic shifts.
We parameterize the results for each model and use a continuous variable $u_{X}$ to smoothly connect between the different models. The MCMC estimation treats the systematics in $X_\mathrm{max}$ using $u_{X}$ as a free parameter uniformly distributed in the interval $[0,1]$ such that
\begin{equation}
\langle X_\mathrm{max}\rangle(u_X) = u_X\langle X_\mathrm{max}\rangle_{QGS} + (1-u_X)\langle X_\mathrm{max}\rangle_{EPOS} 
\end{equation}
\begin{equation}
RMS(X_\mathrm{max})(u_X) = u_X RMS(X_\mathrm{max})_{QGS} + (1-u_X)RMS( X_\mathrm{max})_{EPOS} 
\end{equation}
The parameter $u_{X}$ will be marginalized for the final results. The experimental systematics in the mean $X_\mathrm{max}$ are at the level of 10~g/cm$^{2}$, similar to the systematic differences between the models. Therefore, we do not take them into account.

The likelihood function is Gaussian using the difference of the model and measured values of the mean and RMS of $X_\mathrm{max}$ from~\cite{Porcelli_2015} with their statistical uncertainty as the standard deviation. These measurements subtract the detector effects so they can be compared directly to EAS simulation results.   

\section{Results}
\subsection{Model Parameter Posterior Distributions}
In this section we present the results of our Bayesian inference-based analysis of PAO data and predictions for neutrino fluxes. We evaluate the posterior probability $p(\mathbf{m}|\mathbf{d})$ of the model described in Section 2, with parameter vector $\mathbf{m}$, given the vector of data points $\mathbf{d}$. From Bayes' theorem, the posterior probability is given by 
\begin{equation}
p(\mathbf{m}|\mathbf{d})=\frac{\mathcal{L}(\mathbf{d}|\mathbf{m})\pi(\mathbf{m})}{Z(\mathbf{d})}.
\end{equation}
The likelihood $\mathcal{L}(\mathbf{d}|\mathbf{m})$ is given by the product of the likelihood functions for the energy spectral bin counts of the PAO surface counter data (Section 3.1) and the $\langle X_\mathrm{max}\rangle$ and $RMS(X_\mathrm{max})$ PAO fluorescence data (Section 3.2). The prior probability distribution of model parameters $\pi(\mathbf{m})$ is defined as the product of distributions bounding the parameters (see Table 1 in Section 2.2). As described in Sections 3.1 and 3.2, the systematic uncertainties in the energy scale and models of $X_\mathrm{max}$ are treated as model parameters. Finally, the evidence $Z(\mathbf{d})$ is treated, in this study, as an overall normalization factor. 

To find the set of model parameters $\mathbf{m}$ that is consistent with the data $\mathbf{d}$, we first apply a search that randomly samples an instance of the model vector $\mathbf{m}$ in the full parameter space allowed by $\pi(\mathbf{m})$, including the parameters that account for systematics, and subsequently finds the local minimum. For each trial, we initiate the minimizer using a randomly sampled set of parameters. We perform 60,700 independent randomly-initialized minimizations. The best 5 fits of this run yielded likelihood values ranging from -25.5 to -26.1 with consistent fitted parameters, giving confidence that the global minimum has been identified. The success requirements of this initial step are kept intentionally loose since the objective is to find an initial condition near the global minimum for a more sophisticated analysis. 

The initial set of parameters corresponding to the best fit described above are used as an initial condition for the \texttt{emcee} affine-invariant Markov Chain Monte Carlo~\cite{emcee}. We applied 130 chains (called walkers), each initialized with the best-fit parameters from the initial search independently dithered by a factor of $10^{-8}$. The algorithm proceeds by randomly resampling combinations of parameters from each chain to evaluate $p(\mathbf{m}|\mathbf{d})$ and accepts or rejects the sample based on probability ratios between steps (see~\cite{emcee} for more details). The process of taking a sample $\mathbf{m}$ and the calculation observables (model counts ($N_\mathrm{model}$), $\langle X_\mathrm{max}\rangle$, and $RMS(X_\mathrm{max})$ as a function of energy bin) is referred to as forward modeling. Once convergence is achieved, the ensemble of posterior samples is distributed according to the probability of the model parameters given the data. The ensemble distribution of each single model parameter $m_i$ is the marginalized posterior probability distribution for that parameter.
 
\begin{figure}[!t]
\centering
\includegraphics[width=1.0\linewidth]{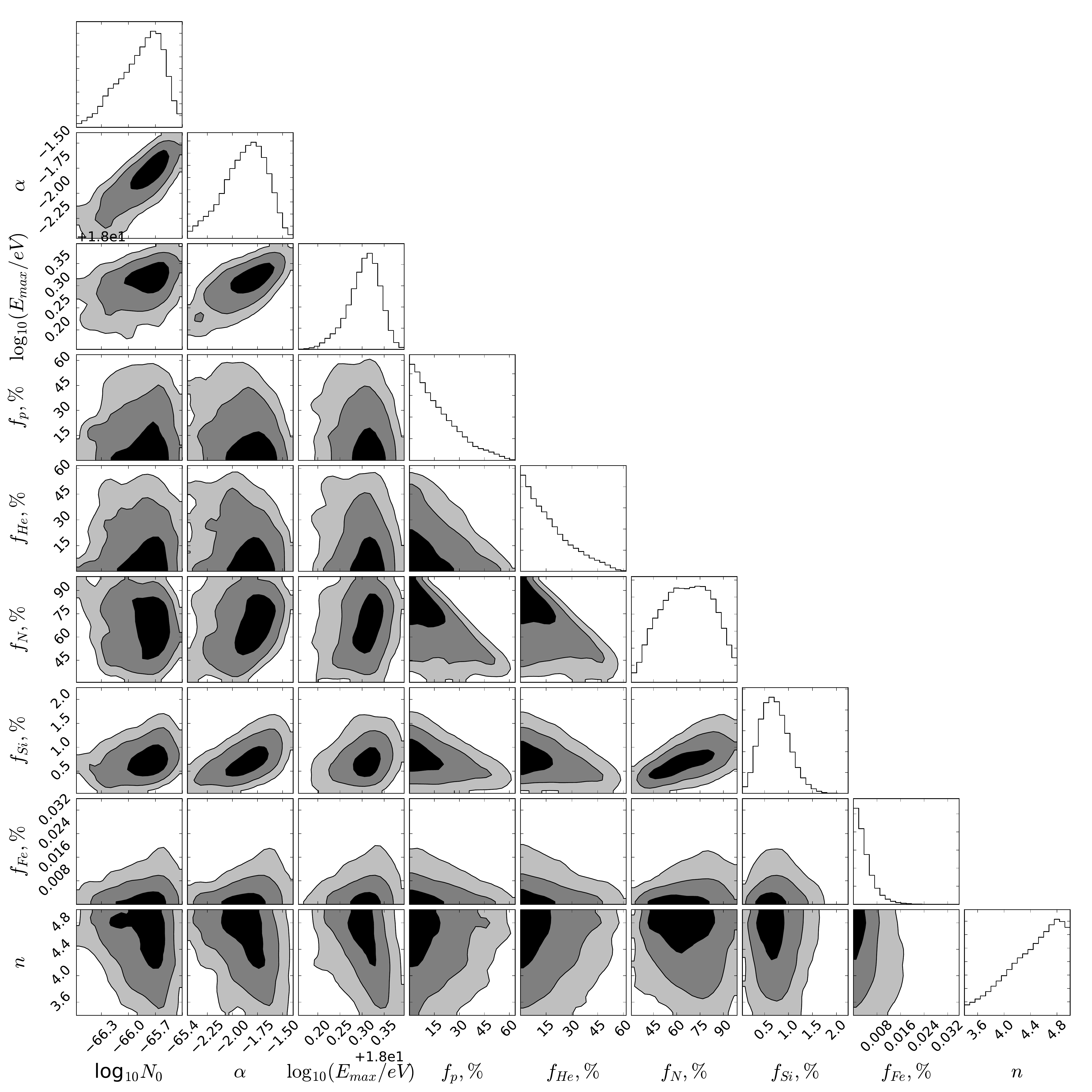}
\caption{Marginalized posterior distributions for normalization, source power law index, maximum energy of acceleration, fractional composition, and source distribution index for a fit assuming astrophysical source distribution. The black, dark and light gray regions correspond to the 1, 2 and 3 $\sigma$ contour levels, respectively. Although not shown here, the results have been marginalized over the source distribution shape parameters ($z_1, z_2, z_\mathrm{max}$) and systematic uncertainties in the energy scale and $X_\mathrm{max}$ models. } 
\label{fig:astroph_parms}
\end{figure}

In Figure~\ref{fig:astroph_parms}, we show the posterior distribution of the normalization $\log_{10}N_0$,  source power law index $\alpha$, maximum energy of acceleration for a proton $E_\mathrm{max}$, the percent fractional composition by element ($f_p$, $f_\mathrm{He}$, $f_\mathrm{N}$, $f_\mathrm{Si}$, $f_\mathrm{Fe}$), and source distribution index. The source distribution shape parameters ($z_1$, $z_2$, $z_\mathrm{max}$) are not shown since they do not provide meaningful marginalizations. The reason is that UHECRs from sources outside the GZK horizon arrive at energies below our cutoff for the likelihood function of $10^{18.9}$~eV. Below this energy, one possible interpretation of the features of the cosmic ray flux is that there is a galactic contribution, which would be negligible above our cutoff.

The source power law index $\alpha$ has 68\% confidence interval $\alpha\in[-2.29, -1.56]$, meaning the source spectrum is a power law rising with energy multiplied by an exponential fall-off (see Equation~\ref{eq:luminosity_model}). Note that this result is only valid for the assumed class of astrophysical sources (Active Galactic Nuclei, Star Formation Rate, and Faranoff-Riley type II galaxies), which are treated as a prior constraint the source evolution parameters (see Table 1). Although this differs significantly from the shock acceleration value of $\alpha\simeq+2$, some of the cases treated by the Auger collaboration in~\cite{PAO_2017} have a spectral index of $\alpha\approx-1.5$, which is consistent with our result, specially considering they restricted the spectral index to values $>-1.5$. For negative values of $\alpha$, the energy at which the spectrum turns from rising to falling is $E=|\alpha| \mathrm{Z}_\mathrm{src} E_\mathrm{max}$, which is, for the posteriors of $\alpha$ obtained here, a factor of $\sim2$ greater than the maximum acceleration energy in the 68\% confidence interval. 

The marginalized posterior distribution of maximum acceleration energy of protons has 68\% confidence interval $\log_{10}\left(E_\mathrm{max} / \mathrm{eV}\right) \in [18.2, 18.4]$. Although this is below our cutoff of $10^{18.9}$~eV for the evaluation of the likelihood function, the spectra for Nitrogen, Silicon, and Iron are well within the energy range tested with Auger data. This value is within the range of energy cutoffs found by~\cite{PAO_2017}, which tested for a variety of propagation models. The results of~\cite{Taylor_2015} are somewhat higher, with $\log_{10}\left(E_\mathrm{max} / \mathrm{eV}\right)$ ranging between 18.6 - 18.9 for $n=+3$. Note, however, that~\cite{Taylor_2015} holds the source evolution index $n$ at fixed value and requires the source spectrum to have a positive spectral index (negative power law), making direct comparison of this the maximum energy cutoff more difficult (see Appendix B).

The proton fraction varies almost entirely through the full range for 0\% up to 60\% in the 95\% confidence interval. The fraction of Helium also ranges between 0\% up to 60\% in the 95\% confidence interval and is anti-correlated with the fraction of protons. The fraction of Nitrogen ranges between 30\% and 100\% in the 95\% confidence interval    
and is the most abundant species of heavy nuclei followed by Silicon contributing $<1.8\%$ and the Iron contribution being negligible. 

The posterior distribution of the source index centered at $n\sim 4.8$ while not strongly discriminating any particular value within the range of priors $n\in[3.4, 5.0]$ consistent with the evolution of astrophysical sources. The remaining source parameters are not shown and do not have meaningful marginalized distributions since UHECR accelerators outside of the GZK horizon result in arrival energies below the energy cutoff used in our likelihood function evaluation. 

\subsection{Comparison to Data and Expected Neutrino Fluence}
\begin{figure}[t!]
\centering
\includegraphics[width=1.0\linewidth]{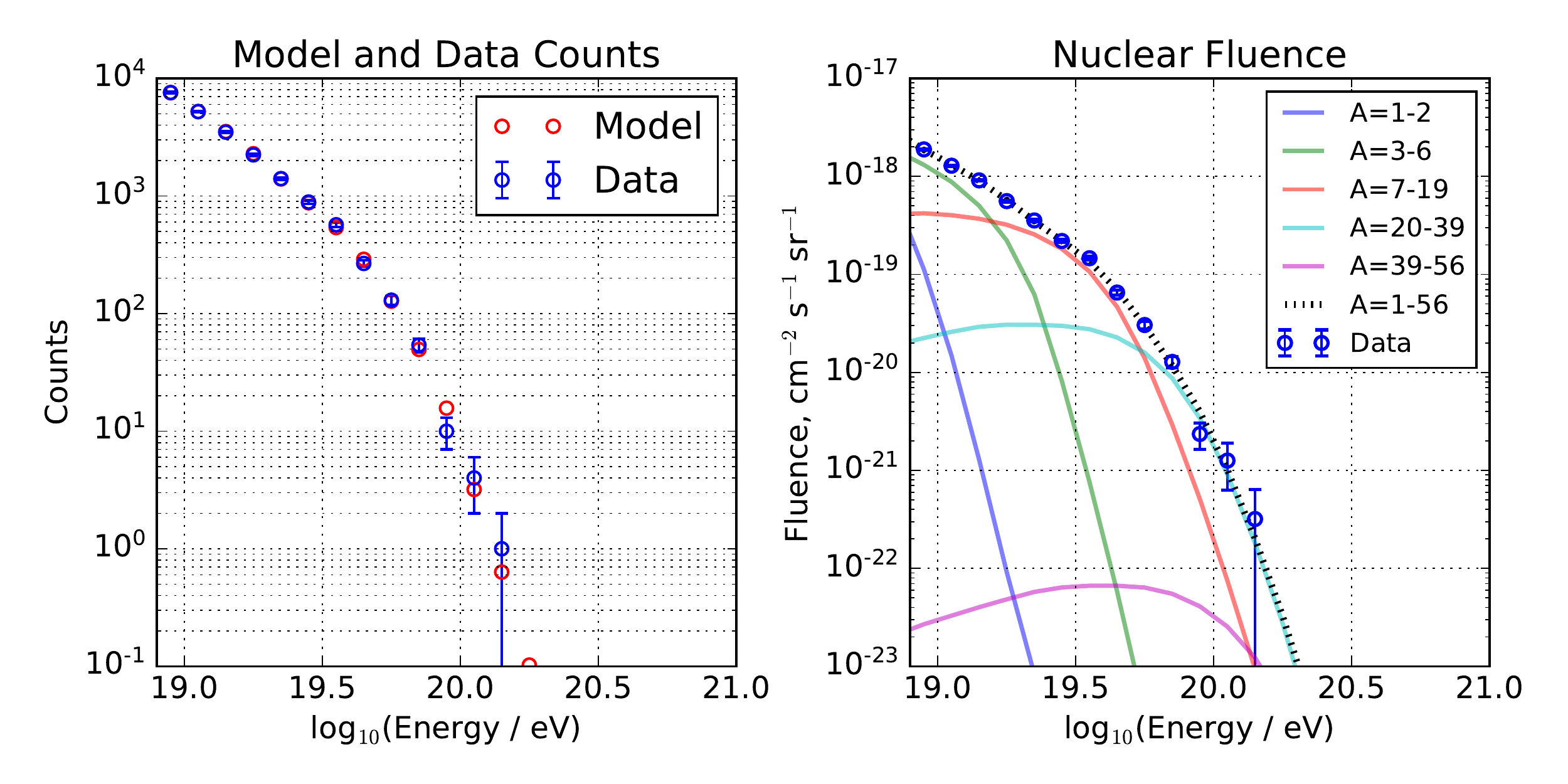}
\caption{Left: Auger particle spectrum counts for the best-fit astrophysical source model compared to data. Right: the contributions to the nuclear fluence are separated into ranges of atomic mass. The best-fit parameters used for this graph have normalization $N_0=1.39\times10^{-66}$, source parameters $\alpha=-2.04$, $\log_{10}(E_\mathrm{max}$)=18.27, source fractional composition $f_\mathrm{p}=45.4\%$, $f_\mathrm{He}=9.3\%$, $f_\mathrm{N}=45.2\%$, $f_\mathrm{Si}=0.05\%$, $f_\mathrm{Fe}=1.5\times 10^{-4}\%$, source evolution index $n=4.8$ with ($z_1, z_2, z_\mathrm{max}$) = (1.45, 3.38, 7.8). The systematic energy shift is -12.3\% with $u_X=0.14$. } 
\label{fig:nuclear_fluence}
\end{figure}

The best-fit nuclear fluence is shown in Figure~\ref{fig:nuclear_fluence}. On the left, we show a comparison of the number of counts expected in each bin for the model and Auger data. Discrepancies are within the 68\% confidence Poisson error bars shown in the plot. On the right panel of Figure~\ref{fig:nuclear_fluence} we show the fluence resulting from the best-fit model compared to the data and show the contribution for range of atomic masses of A between 1-2, 3-6, 7-19, 20-39, and 39-56. For energies $>10^{18.9}$~eV, the range of elements with atomic mass 3-6, 7-19, 20-39 dominate the spectrum in different energy regions. The best-fit values are provided in the caption of Figure~\ref{fig:nuclear_fluence}.

\begin{figure}[h!]
\centering
\includegraphics[width=0.9\linewidth]{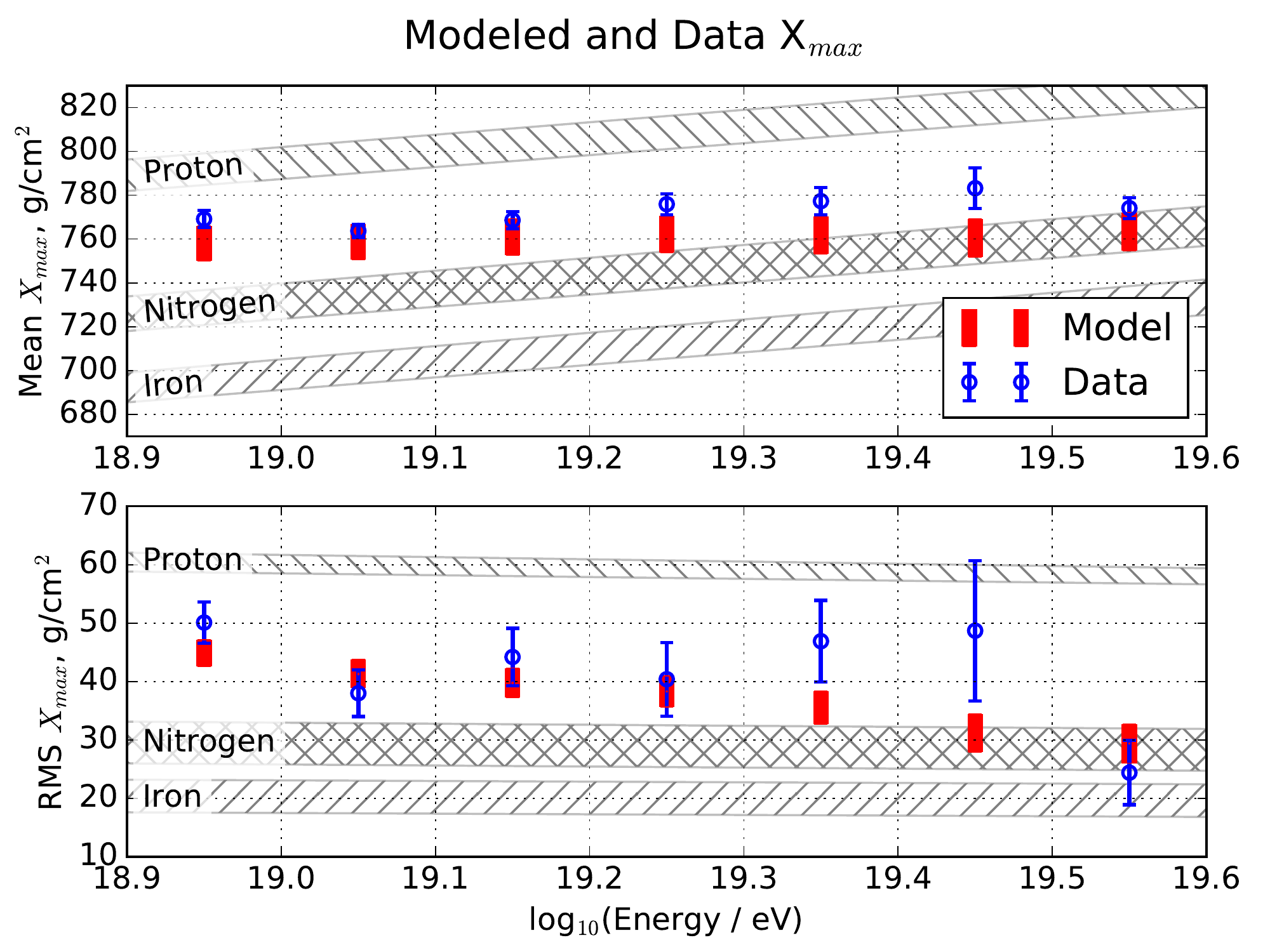}
\caption{Top: the mean value of $X_\mathrm{max}$ for Auger data and the best-fit model as a function of energy. Bottom: the root-mean-square (RMS) value of $X_\mathrm{max}$ for Auger data and the best fit model as a function of energy. The error bars for the model indicate the range between the QGSJet and EPOS-LHC models. The hatched regions show the $\langle X_\mathrm{max}\rangle$ and $RMS(X_\mathrm{max})$ for pure proton, Nitrogen, and Iron nuclei.} 
\label{fig:Xmax}
\end{figure}

The best-fit $X_\mathrm{max}$ mean and RMS values are shown in Figure~\ref{fig:Xmax}. 
The model range corresponds to the systematic differences between the EPOS-LHC and QGSJetII-04 hadronic models.
The model mean $X_\mathrm{max}$ shows a slight overall discrepancy with the data of $\sim 5$ g/cm$^{2}$ but it is consistent with the systematic uncertainty of the measurements.
The data and best-fit model for the RMS $X_\mathrm{max}$ are in good agreement. 

\begin{figure}[h!]
\centering
\includegraphics[width=0.8\linewidth]{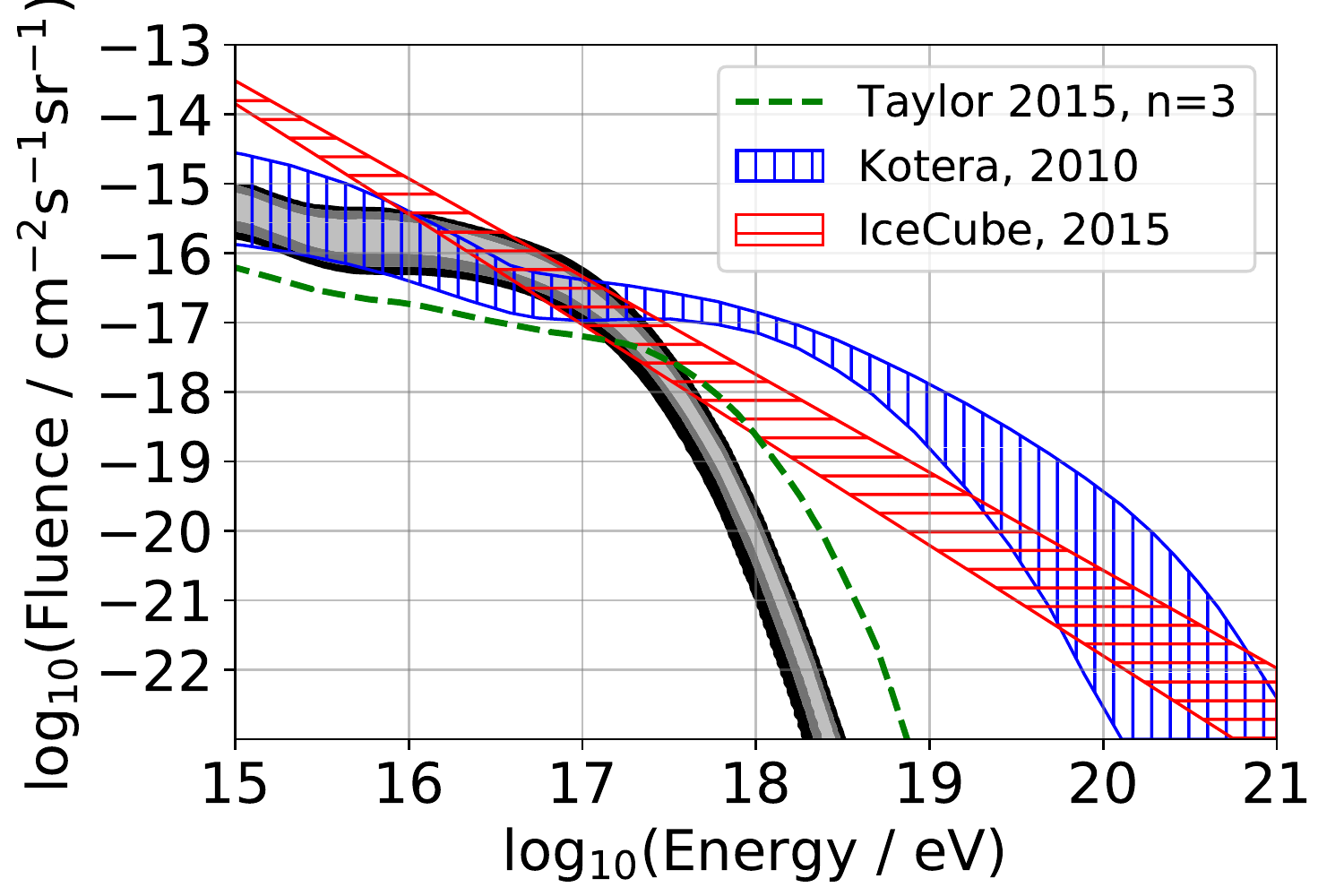}
\caption{Neutrino fluence estimates based on posterior distributions of the astrophysical model parameters presented in this study. The lightest gray band corresponds to 68\% of parameters populating each energy value. The darker regions correspond to the 95\% and 99\% of parameters populating each energy bin. The blue curves correspond to the bounds of the range of models from Kotera 2010~\cite{Kotera_2010}. The dashed red lines bound the 68\% confidence interval of the IceCube fitted parameters for a single power law~\cite{IceCube_2015}. Here we have extrapolated the curves beyond the highest IceCube energy bin with 1-2 PeV. The results from Taylor 2015~\cite{Taylor_2015} for source evolution index $n=+3$ are shown in the green dashed line.} 
\label{fig:nu_fluence}
\end{figure}

The resulting neutrino fluence using the posterior distributions are shown in Figure~\ref{fig:nu_fluence}. The gray shaded regions in Figure~\ref{fig:nu_fluence} represent how often a set of posterior parameters lands in a given observed energy and fluence. The light to dark shades indicate the 68\%, 95\%, and 99\% occurrence. 

In Figure~\ref{fig:nu_fluence}, the curves bounding a range of models from Kotera 2010~\cite{Kotera_2010} are shown in blue with vertical-line hatch pattern. The stark difference between the results of this study and the Kotera 2010 predictions is due to $E_{max}$. For the results in this study, the suppression of the flux at ($E>10^{18}$~eV) is primarily due to the maximum acceleration energy posteriors 68\% confidence interval of $\log_{10}\left(E_\mathrm{max} / \mathrm{eV}\right)\in[18.2, 18.4]$, which is lower than the energies considered in the range of Kotera 2010 models shown.

The extrapolation of the extra-terrestrial neutrino flux reported by the IceCube experiment~\cite{IceCube_2015} is also shown in Figure~\ref{fig:nu_fluence}. This flux is likely not cosmogenic and its overlap with our prediction for {\it cosmogenic} neutrinos may pose problems in the future to disentangle both components. On the other hand, it can be said that the experimental sensitivity required to prove the IceCube flux at higher energies is comparable to the one required to detect {\it cosmogenic} neutrinos, which further motivates experimental efforts.
The curves extend well beyond the PeV-scale maximum energies reported to compare the potential overlap between IceCube and cosmogenic neutrino fluxes. For 10-100 PeV energies, the fluxes are comparable. If the PeV flux of neutrinos discovered by IceCube were to have no cutoff below $10^{18}$~eV, the results of this study indicate they may dominate over the cosmogenic flux at ultra-high energies. 

The results from Taylor 2015~\cite{Taylor_2015} for source evolution index $n=+3$ are also shown in Figure~\ref{fig:nu_fluence}. The predicted flux from this study is lower than our results for $\log_{10}(E/\mathrm{eV})<17.5$ and higher for $\log_{10}(E/\mathrm{eV})>17.5$. Direct comparison is difficult due to the different approaches taken. However, part of the reason for the discrepancy at lower energies is likely due to fixing the source evolution index $n=+3$ in contrast to our approach where $n\in[3.4,5.0]$. This results in a larger number of sources that contribute to the neutrino flux at redshifts that do not contribute to the UHECR flux at energies $E\geq10^{18.9}$~eV. The best-fit value for the spectral index for Taylor 2015 with $n=+3$ is $\alpha=0.64$, which is significantly different from our best-fit value of $\alpha=-2.04$. It is also worth noting that the source-flux parameterizations are different. At high energies, the equivalent of Taylor 2015 best-fit results for $\log_{10}(E_\mathrm{max}/\mathrm{eV})=18.75$, which is higher than our best-fit $\log_{10}(E_\mathrm{max}/\mathrm{eV})=18.27$. The larger value of $E_\mathrm{max}$ is expected to produce a higher neutrino fluence at higher energies.

\subsection{Sensitivity of Current and Proposed Neutrino Experiments}
\begin{figure}[h!]
\centering
\includegraphics[width=0.49\linewidth]{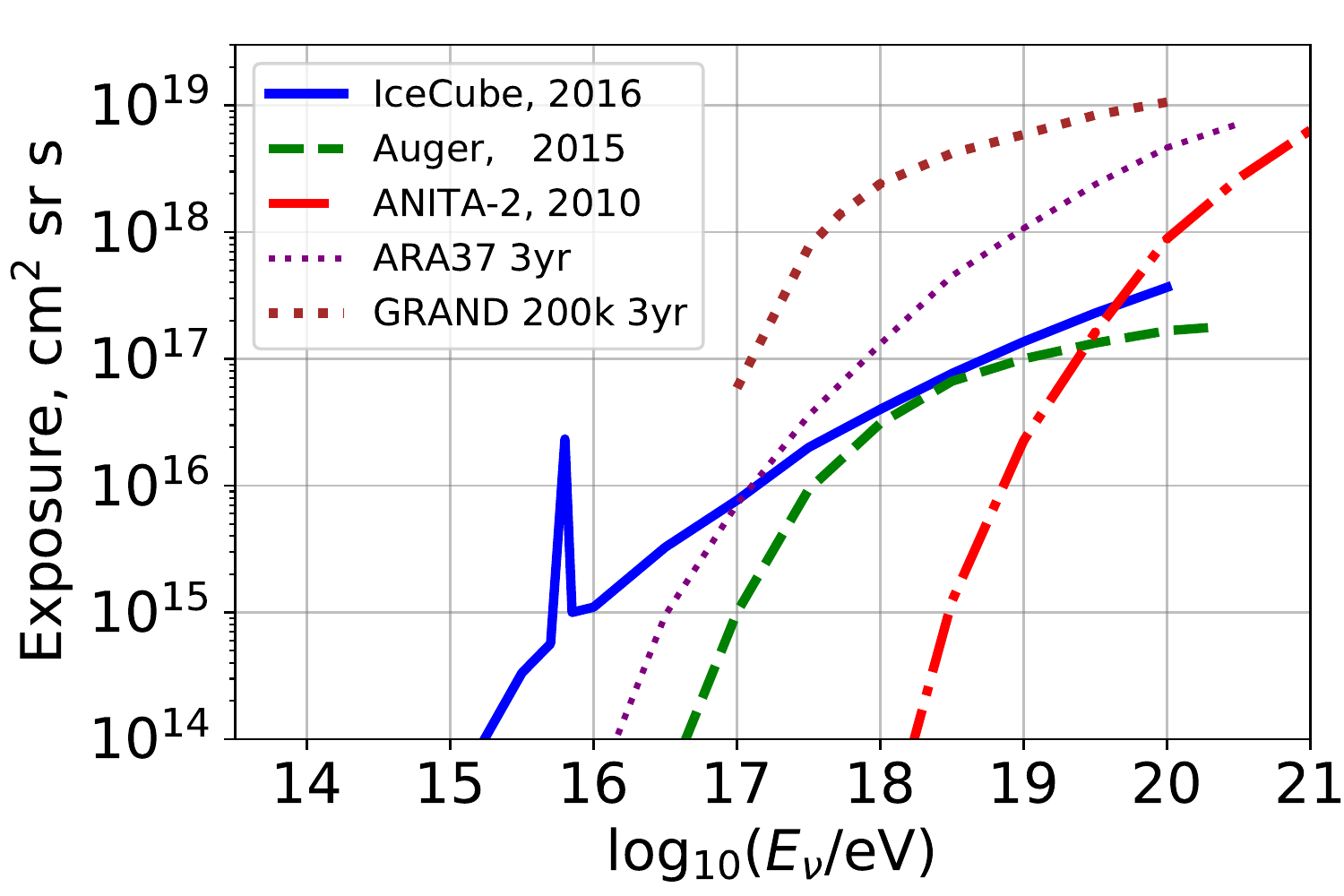}
\includegraphics[width=0.49\linewidth]{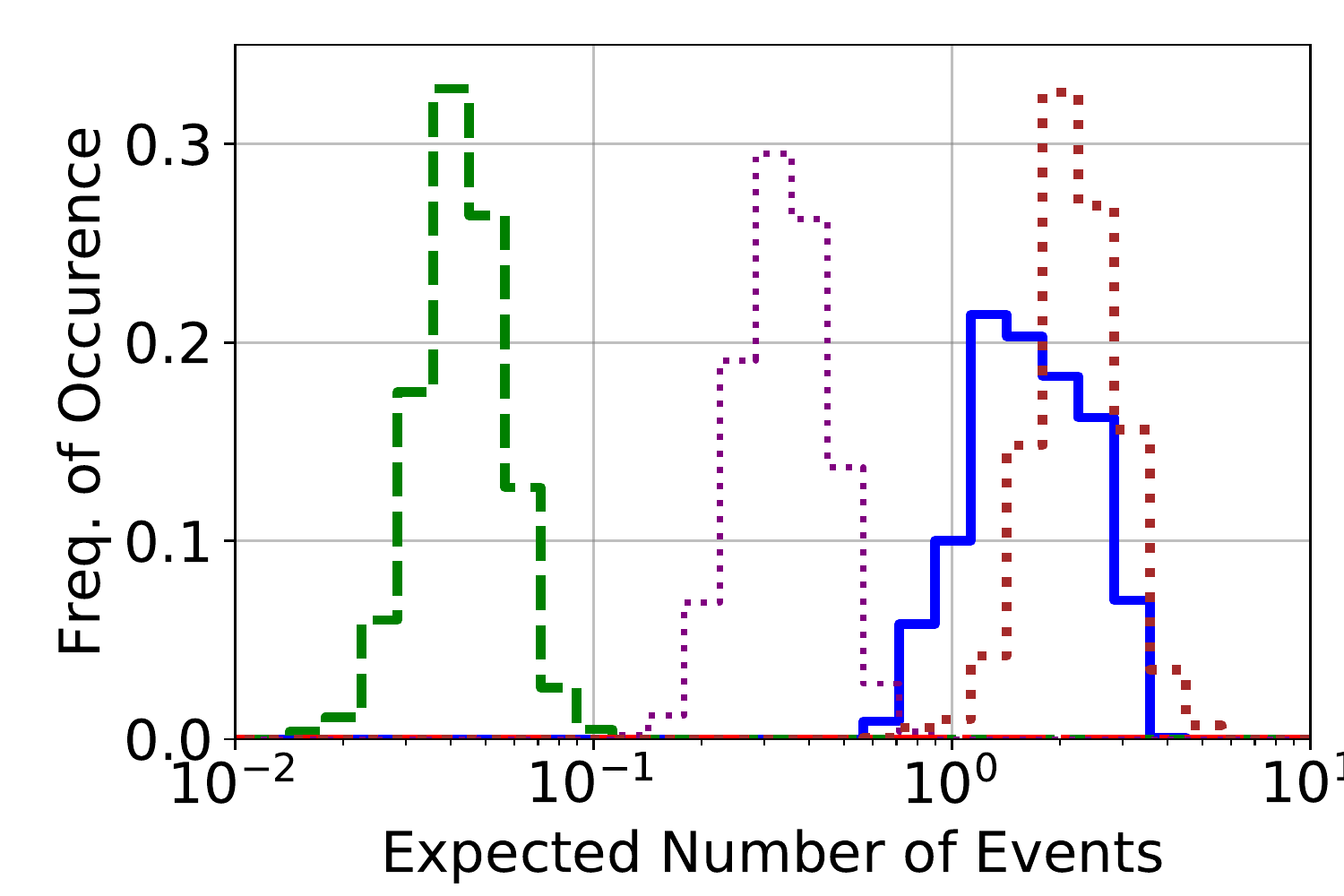}
\caption{Exposure and expected number of events for various current and proposed experiments. The exposures presented assume a $\nu_e$:$\nu_\mu$:$\nu_\tau$=1:1:1 flavor ratio on Earth. On the left panel, the solid blue line corresponds to the IceCube ultra-high energy search~\cite{IceCube_UHE_2016} flavor-averaged exposure, the dashed green line corresponds to the Auger~\cite{Auger_Neutrino_2015} $\nu_\tau$ exposure, and the dash-dotted red line corresponds to the ANITA-2~\cite{Gorham_2010} flavor-averaged exposure. For proposed experiments, the dotted purple line corresponds to the ARA 37-station flavor-averaged 3-year exposure~\cite{ARA_2016} and the thicker brown dotted line corresponds to the GRAND~\cite{GRAND_2017} $\nu_\tau$ 200k antenna 3-year exposure. In the right panel we show the distribution of expected number of events corresponding to each experiment (line styles and colors are the same as on the left panel) and the distribution of neutrino fluence produced in this work.} 
\label{fig:exposure_counts}
\end{figure}

We estimate the sensitivity of current and future detectors to the fluxes presented here by estimating the range of expected event rates. For an exposure as a function of energy $\mathcal{E}(E)$ and fluence $EF(E)$, the expected number of events is given by
\begin{equation}
\langle N \rangle = \ln(10)\int_{E_\mathrm{min}}^{E_\mathrm{max}} d\log_{10}(E)  \ \mathcal{E}(E) \  EF(E).
\end{equation}
We sample the posterior distribution of parameters to produce neutrino flux curves, and estimate the corresponding distribution of events for current and proposed instrument exposures. In the left panel of Figure~\ref{fig:exposure_counts} we show exposures assuming a $\nu_e$:$\nu_\mu$:$\nu_\tau$=1:1:1 flavor ratio on Earth. Namely, the IceCube ultra-high energy search~\cite{IceCube_UHE_2016} flavor-averaged exposure, the Auger~\cite{Auger_Neutrino_2015} $\nu_\tau$ exposure, and the ANITA-2~\cite{Gorham_2010} flavor-averaged exposure. For proposed experiments, we include the ARA 37-station flavor-averaged 3-year exposure~\cite{ARA_2016} and the GRAND~\cite{GRAND_2017} $\nu_\tau$ 200k antenna 3-year exposure.

On the right of Figure~\ref{fig:exposure_counts} we show the expected number of events for these experiments for the posterior neutrino fluence estimated in this work (see Figure~\ref{fig:nu_fluence}). The distributions of expected number of events are purely due to the distribution of posteriors from the fit to PAO data and do not include additional uncertainties due to each neutrino detector. The expected number of events for ANITA-2 (not shown) ranges between $10^{-8}$ and $10^{-7}$. The Auger experiment would expect $\sim0.05$ events. Both limits from ANITA-2 and Auger are consistent with the cosmogenic neutrino fluence presented here. The IceCube ultra-high energy search is expected to have yielded between 1.2~--~2.7 events at energies $>10^{15}$~eV. This sensitivity is potentially enough to place additional constraints on the astrophysical source model presented here, which will be the subject of a follow-up paper. 

The expected number of events for proposed experiments are shown on the right panel of Figure~\ref{fig:exposure_counts}. ARA37 is expected to produce $\sim0.4$ events, which would not significantly constrain the neutrino fluence results presented here. The GRAND 200k antenna array is expected to obtain between 1.8~--~3.3 events, which is comparable to IceCube. The advantage of GRAND is that the events would be at energies $>10^{17}$~eV, which is in a different energy band from IceCube. The combination of GRAND and IceCube results has the potential to place significant constraints on neutrino production models in general. 


\section{Conclusions}

We have presented a Bayesian inference-based estimate of the cosmogenic neutrino flux on Earth that uses the current UHECR measurements of the Pierre Auger Observatory as a constraint.
The derived neutrino fluxes are robust as long as these conditions are fulfilled: the density of sources per Mpc$^{3}$ is large enough so \refeq{FluxModel} applies, the source redshift distribution resembles the evolution of some typical astrophysical sources (Active Galactic Nuclei, Star Formation Rate, and Faranoff-Riley type II galaxies), interaction losses at the source are small so the cutoff of the spectrum depends on rigidity, the source spectrum and nuclear composition do not change significantly with redshift, the cosmic background photon fields and cross sections are correctly modeled, and the effect of IGMF is small. 
Within this context, the Bayesian inference method provides the full statistical range of neutrino fluxes. By using priors in astrophysical source distributions, including the uncertainty associated with source density at high redshift, the method takes into account contributions to the neutrino flux from distant sources that do not contribute to the observed UHECR flux. 

Under these assumptions, we find that the cosmogenic flux of ultra-high energy neutrinos ($E_{\nu}\gtrsim10^{18}$~eV) is strongly suppressed compared to the models of Kotera~2010~\cite{Kotera_2010} due to the lower value of the maximum acceleration energy of protons ($E_\mathrm{max}\sim18.3$). Within the variety of model choices and systematics, the studies of the Auger collaboration~\cite{PAO_2017}~and~Taylor~2015~\cite{Taylor_2015} have found comparably low values of $E_\mathrm{max}$, which has a dominant effect on the production of cosmogenic neutrinos at the highest energies: the maximum acceleration is below the threshold of neutrino production by protons interacting with the CMB at redshift $z=0$. 

We tested the sensitivity of current and proposed experiments on the neutrino fluence resulting from this work. It is possible that a fraction of IceCube events at energies $>10^{15}$~eV are due to cosmogenic neutrinos. The impact of the IceCube results on the model presented here will be the subject of a future study. Currently, GRAND is the only proposed experiment that would have sensitivity to the neutrino fluence resulting from this work and could provide additional independent constraints to the model presented here in a higher energy band ($>10^{17}$~eV) than IceCube.

It is important to highlight the assumption used in this study that sources of ultra-high energy cosmic rays behave the same independent of redshift. The ultra-high energy cosmic-ray flux observed on Earth is dominated by sources within several hundred Mpc while it is expected that the neutrino flux at the highest energies is dominated by sources significantly farther away. The discovery of a significant flux of neutrinos at ultra-high energies could be due to a change in the acceleration characteristics of sources at distances away from the GZK horizon. It could, however, also be due to an unbroken spectrum of the neutrino flux discovered by IceCube or perhaps even by neutrinos produced within the astrophysical sources themselves. 

On the basis of the method presented here, we will investigate in upcoming publications the four main effects that can significantly change the neutrino flux levels: 1) a break of the rigidity ansatz (where the energy cutoff of each element is proportional to the nuclear charge and the maximum energy of the proton), 2) scenarios in which the observed UHECR flux is due a few nearby sources, 3) the effect of the uncertainties of extra-galactic magnetic field, and 4) uncertainties in the cosmic background photon fields and cross sections. 
It is important to note that interaction losses at the source (break up of the rigidity ansatz) usually lead to production of neutrinos that could be tagged by correlation with astrophysical catalogs; {\it cosmogenic} neutrinos are probably diffuse due to UHECR propagation in turbulent IGMF. 
The second effect is related with the cosmic variance: if the source density is small and we happen to be close to an accelerator, the UHECR flux would be enhanced compared to the average. If this is the case, the observed flux will resemble the source flux and, on the basis of the source characteristics and astrophysical catalogs, a baseline prediction of the {\it cosmogenic} flux could still be produced. 
For the third effect, the uncertainties in the IGMF can modify the cosmic ray and neutrino fluxes. Although the effects on the cosmic-ray flux have been studied, the impact on the cosmogenic neutrino flux has not been characterized in detail.
Finally, the fourth effect is important and cannot be disregarded. Although the effect has been studied for the cosmic ray flux~\cite{Alves_Batista_2015}, it has not been done for neutrinos. The cosmogenic neutrino flux levels presented in the energy band 10-100 PeV are comparable with an extrapolation of the reported IceCube extra-terrestrial neutrino flux. Experiments targeting this energy range could potentially detect contributions from two neutrino populations, increasing the odds of observing events. An exposure at least a factor of 10 better than IceCube would be required. 

\

\noindent{\it Acknowledgements:} Part of this work was carried out at the Jet Propulsion Laboratory, California Institute of Technology, under a contract with the National Aeronautics and Space Administration. This work has been supported by a research grant from the Conselho Nacional de Desenvolvimento Científico e Tecnológico (CNPq). We also thank Prof. V. de Souza Filho and the Instituto de Fisica de São Carlos (USP) for their continuous support. \textsf{\copyright} 2017. All rights reserved.

\appendix

\section{Derivation of Flux Formula}

Suppose a source at redshift $z$ accelerates a nucleus of charge Z$_\mathrm{src}$ and mass A$_\mathrm{src}$ to energy $E_\mathrm{src}$. As the nucleus propagates to redshift $z=0$, it can fragment into nuclei of mass A, neutrinos, and photons. The spectral yield of the number observed secondaries $N_{k,\mathrm{obs}}$ of species $k$ at energies between $E_\mathrm{obs}$ and $E_\mathrm{obs} + dE_\mathrm{obs}$ per nucleus of species (Z$_\mathrm{src}$, A$_\mathrm{src}$) per number of source particles $N_\mathrm{src}$ at redshift $z$ with energy $E_\mathrm{src}$  is expressed as
\begin{equation}
{Y}_{k}(E_\mathrm{obs} | E_\mathrm{src}, z, (\mbox{Z}_\mathrm{src}, \mbox{A}_\mathrm{src}))
=
\frac{d^2N_{k,\mathrm{obs}}(E_\mathrm{obs} | E_\mathrm{src}, z, (\mbox{Z}_\mathrm{src}, \mbox{A}_\mathrm{src}))}{dN_\mathrm{src} dE_\mathrm{obs}}.
\end{equation} 

The value of this function is obtained via propagation simulations stepped in redshift and source energy for various source nuclear species. The observed energy spectrum of a given species $k$ at energy $E_\mathrm{obs}$ is the result of all primary species and source energies whose decay products result in such secondary particles. 
This is expressed via the integral summed over source nuclear species
\begin{equation}
\begin{split}
\frac{d^2N_{k,obs}(E_\mathrm{obs} | z)}{dN_\mathrm{src}dE_\mathrm{obs}} 
= \sum_{\left(\mbox{Z}_\mathrm{src}, \mbox{A}_\mathrm{src}\right)} 
\ \int dE_\mathrm{src} &
\ {Y}_{k}(E_\mathrm{obs} | E_\mathrm{src}, z, (\mbox{Z}_\mathrm{src}, \mbox{A}_\mathrm{src}))
\\
& \ \times \frac{dN_\mathrm{src}(E_\mathrm{src}|z, (\mbox{Z}_\mathrm{src}, \mbox{A}_\mathrm{src}))}{dE_\mathrm{src}}.
\label{eqn:yield_relation}
\end{split}
\end{equation} 

The flux of particles of species $k$ due to a single source at redshift $z$ is expressed as a differential rate according to
\begin{equation}
I_{k,\mathrm{obs}}(E_\mathrm{obs}|z) = \frac{d^4N_{k,\mathrm{obs}}(E_\mathrm{obs}|z)}{dE_\mathrm{obs}dt_\mathrm{obs}d^2G_\mathrm{obs}}
\end{equation}
Differentiating Equation~\ref{eqn:yield_relation} with respect to time $t_\mathrm{obs}$ and \'etendu $G_\mathrm{obs}$ (or acceptance) and applying the chain rule gives
\begin{equation}
\begin{split}
I_{k,obs}(E_\mathrm{obs}|z) = 
 \sum_{\left(\mbox{Z}_\mathrm{src}, \mbox{A}_\mathrm{src}\right)}
\ \int dE_\mathrm{src} &
\ {Y}_{k}(E_\mathrm{obs} | E_\mathrm{src}, z, (\mbox{Z}_\mathrm{src}, \mbox{A}_\mathrm{src})) \\
& \times \frac{d^4N_\mathrm{src}(E_\mathrm{src}|z,(\mbox{Z}_\mathrm{src}, \mbox{A}_\mathrm{src}))}{dE_\mathrm{src}dt_\mathrm{src}d^2G_\mathrm{src}}
\frac{dt_\mathrm{src}}{dt_\mathrm{obs}}
\frac{d^2G_\mathrm{src}}{d^2G_\mathrm{obs}}
\end{split}
\label{eq:integral_1}
\end{equation}

The source flux is
\begin{equation}
I_\mathrm{src}(E_\mathrm{src}|z, (\mbox{Z}_\mathrm{src}, \mbox{A}_\mathrm{src})) = \frac{d^4N_\mathrm{src}(E_\mathrm{src}|z,(\mbox{Z}_\mathrm{src}, \mbox{A}_\mathrm{src}))}{dE_\mathrm{src}dt_\mathrm{src}d^2G_\mathrm{src}}
\end{equation}

The derivatives in the integrand of Equation~\ref{eq:integral_1} are determined by the the following relations. The \'etendue is
\begin{equation}
(1+z)^2 d^2G_\mathrm{src} = d^2G_\mathrm{obs}.
\end{equation}
The observed rate of emission will be retarded by the redshift factor, which gives
\begin{equation}
dt_\mathrm{src} = (1+z)dt_\mathrm{obs}.
\end{equation}
This results in 
\begin{equation}
I_{k,obs}(E_\mathrm{obs}|z) = 
 \sum_{\left(\mbox{Z}_\mathrm{src}, \mbox{A}_\mathrm{src}\right)}
\ \int  
\ dE_\mathrm{src} \
{Y}_{k}(E_\mathrm{obs} | E_\mathrm{src}, z, (\mbox{Z}_\mathrm{src}, \mbox{A}_\mathrm{src})) \ \frac{I_\mathrm{src}(E_\mathrm{src}|z, (\mbox{Z}_\mathrm{src}, \mbox{A}_\mathrm{src}))}{(1+z)}
\end{equation}

For an isotropic emitter with a cosmic-ray Luminosity per energy band 
\begin{equation}
L_\mathrm{src}=\frac{d^2N_\mathrm{src}}{dt_\mathrm{src} dE_\mathrm{src}}
\end{equation}
is related to the flux via
\begin{equation}
L_{\mathrm{Z}_\mathrm{src}, \mathrm{A}_\mathrm{src}}(E_\mathrm{src}) = 4\pi d_C^2 I_{\mathrm{Z}_\mathrm{src}, \mathrm{A}_\mathrm{src}}(E_\mathrm{src}).
\end{equation}
where $d_C$ is the comoving distance. Applying this relation results in
\begin{equation}
\begin{split}
I_{k,obs}(E_\mathrm{obs}|z) = 
 \sum_{\left(\mbox{Z}_\mathrm{src}, \mbox{A}_\mathrm{src}\right)}
\ \int 
\ dE_\mathrm{src}   \ 
{Y}_{k}(E_\mathrm{obs} | E_\mathrm{src}, z, (\mbox{Z}_\mathrm{src}, \mbox{A}_\mathrm{src})) \ \frac{L_\mathrm{src}(E_\mathrm{src}|z, (\mbox{Z}_\mathrm{src}, \mbox{A}_\mathrm{src}))}{(1+z)4\pi d_C^2}
\end{split}
\end{equation}

We want to integrate the flux over all sources in a shell between redshift $z$ and $z+dz$. Suppose the density of sources per comoving volume is given by $n_\mathrm{src}(z)=dN_\mathrm{src}/dV_C$. For a flat universe, the differential comoving volume is given by
\begin{equation}
dV_C = \frac{c}{H_0}\frac{d_C^2}{\mathcal{H}(z)}d\Omega dz
\end{equation}
where $\mathcal{H}(z)=\sqrt{\Omega_M(1+z)^3+\Omega_{\Lambda}}$. The differential volume of a spherical shell between redshift $z$ and $z+dz$ is
\begin{equation}
dV_{shell} = \frac{c}{H_0}\frac{4\pi d_C^2}{\mathcal{H}(z)}dz.
\end{equation}
The total number of sources between redshift $z$ and $z+dz$ is given by $n_\mathrm{src} dV_{shell}(z)$

\begin{equation}
\begin{split}
I_{k,obs}(E_\mathrm{obs}) = 
 \sum_{\left(\mbox{Z}_\mathrm{src}, \mbox{A}_\mathrm{src}\right)}
\ \int n_\mathrm{src}(z) \ dV_{shell} 
\ \int dE_\mathrm{src} \ &
{Y}_{k}(E_\mathrm{obs} | E_\mathrm{src}, z, (\mbox{Z}_\mathrm{src}, \mbox{A}_\mathrm{src})) \\ 
& \times \frac{L_\mathrm{src}(E_\mathrm{src}|z, (\mbox{Z}_\mathrm{src}, \mbox{A}_\mathrm{src}))}{(1+z)4\pi d_C^2}
\end{split}
\end{equation}

The total flux of particles of species $k$ is given by
\begin{equation}
\begin{split}
I_{k}(E_\mathrm{obs}) & = \sum_{(\mathrm{Z}_\mathrm{src}, \mathrm{A}_\mathrm{src})}
\int dz \ \frac{c}{H_0}\frac{n_\mathrm{src}(z)}{(1+z)\sqrt{\Omega_M(1+z)^3+\Omega_{\Lambda}}}  \\
& \ \ \ \ \ \ \ \ \ \ \ \ \ \ \  \times \int dE_\mathrm{src} \  
{Y}_{k}(E_\mathrm{obs} | E_\mathrm{src}, z, \mbox{Z}_\mathrm{src}, \mbox{A}_\mathrm{src}) \ L_{\mathrm{Z}_\mathrm{src},\mathrm{A}_\mathrm{src}}(E_\mathrm{src}|z)
\end{split}
\end{equation} 

In practice, the yield function is calculated in logarithmic bins $d\log_{10}E_\mathrm{obs}$ and the redshifts and energies are integrated logarithmically. A change of variables gives $dx = \ln(10) \ x \ d\log_{10}(x)$. Define the yield function logarithmically binned in $d\log_{10}E_\mathrm{obs}$ as $\hat{Y}_k$, which is related to $Y_k$ via
\begin{equation}
{\hat{Y}}_{k}(E_\mathrm{obs} | E_\mathrm{src}, z, \mbox{Z}_\mathrm{src}, \mbox{A}_\mathrm{src})
=
\frac{1}{\ln(10)E_\mathrm{obs}}
Y_{k}(E_\mathrm{obs} | E_\mathrm{src}, z, \mbox{Z}_\mathrm{src}, \mbox{A}_\mathrm{src})
\end{equation}
Applying this change of variables to $z$ and $E_\mathrm{src}$, the integral is approximated numerically as
\begin{equation}
\begin{split}
I_{k}(E_\mathrm{obs})  = \sum_{(\mathrm{Z}_\mathrm{src}, \mathrm{A}_\mathrm{src})}\sum_i\sum_j 
& \ \ln(10)\frac{c}{H_0}\frac{ \Delta \log_{10}(z)\Delta \log_{10}(E_\mathrm{src} / eV) }{(1+z_i)\sqrt{\Omega_M(1+z_i)^3+\Omega_{\Lambda}}} 
 \    \frac{E_\mathrm{src,j}}{E_\mathrm{obs}} \ z_i  \\
& \times n_\mathrm{src}(z_i) \ {\hat{Y}}_{k}(E_\mathrm{obs} | E_\mathrm{src,j}, z, \mbox{Z}_\mathrm{src}, \mbox{A}_\mathrm{src}) \ L_{\mathrm{Z}_\mathrm{src},\mathrm{A}_\mathrm{src}}(E_\mathrm{src,j}|z)
\end{split}
\end{equation} 

The units, constants, and scales are:
\begin{itemize}
\item{$H_0=67.6$~km/s/Mpc = $(4.56\times10^{17})^{-1}$ s$^{-1}$}
\item{$c=2.998\times10^{10}$~cm/s}
\item{$c/H_0=1.37\times 10^{28}$~cm}
\item{$n_{0,src}\sim 10^{-3}~\mbox{Mpc}^{-3} = 3.4 \times 10^{-77} \mbox{cm}^{-3}$ }
\item{$L_{0,src}\sim 10^{12}~\mbox{eV}^{-1}~\mbox{s}^{-1}$}
\end{itemize}

\section{Comparison to Previous Results}
The approach in this paper is different from previous similar studies by the Auger collaboration~\cite{PAO_2017} (referred to as PAO17 in this section) and Taylor {\it et al.}~2015~\cite{Taylor_2015} (referred to as TAH15) in that the source evolution has been given a functional shape consistent with Gamma Ray Bursts, Active Galactic Nuclei, Star Formation Rate, and Faranoff-Riley type II galaxies as described in~\cite{Kotera_2010}. To compare with PAO17 and TAH15 we have produced a simulation run that sets the source evolution index to the fixed value $n=0$ and limits the contribution to sources with $z\leq0.5$, which is the set of parameters most commonly used in the variety of scenarios studied in PAO17. 

Table~\ref{tab:comparison} lists the results for maximum acceleration energy, source spectral index, and fractional source composition. We have labeled this work RA18 and we report the shortest 68\% confidence intervals resulting from our fit with source index $n=0$ and redshift cutoff $z=0.5$. We have treated the systematics as nuisance parameters as discussed in the main text of this paper. For the results of TAH15, we have adapted their reported values to an interval for ease of comparison. The range of fitted values is in general agreement except for the source spectral index, which our result gives as $\alpha\sim0$ while TAH15 has $\alpha\sim+1.6$.

There are several results reported in PAO17 covering a range of models and fitting approaches. Their reference model (labeled SPG in Table~\ref{tab:comparison}) uses the SimProp propagation code with Puget-Stecker-Bredekamp photo-disintegration model and the Gilmore 2012 extragalactic background light model along with the EPOS-LHC hadronic model (see PAO17 for references). The values in Table~\ref{tab:comparison} are the set labeled ``main minimum average" values from Table~1 in PAO17, presented here as an interval for ease comparison. Note that these values do not include systematic uncertainties. The row labeled ``w/ sys" are the values including systematic uncertainties as nuisance parameters reported in~PAO17. Significant differences appear when systematic uncertainties are included.

The closest approach treated in~PAO17 to the one implemented in this study is the ``CTG" model using the CRPropa propagation model with TALYS photo-disintegration model and the Gilmore 2012 extragalactic background light model along with the EPOS-LHC hadronic model (see~PAO17 for references). They do not include the effect of systematic uncertainties in the values reported in this row. They report two values corresponding to comparable minima in their fit, which we label CTG$^{1}$ and CTG$^{2}$ corresponding to the first and second minima. CTG$^1$ gives negative source index value, with $\alpha\sim-1$, while the other fits shown here give positive values with $\alpha\sim+1$ while our fit results in $\alpha\sim0$. The maximum acceleration energy resulting from our study is in general agreement with TAH15 and the range of results in PAO17.

There are many more variants of the source index parameter $n=0$ reported in PAO17~\cite{PAO_2017} that produce source spectral indices ranging from $\alpha=-1.5$ (values below which they do not consider) and $\alpha=+2.1$ depending on the combination of propagation model, photo-disintegration cross section, extragalactic background light model, and hadronic model in addition to the inclusion of systematic uncertainties as nuisance parameters. Tables 5, 6, and 8 of PAO17 show the sign and value of the source spectral index are not stable for various model inputs. No one combination reported in~PAO17 precisely matches the combination of models and use of nuisance parameters treated here. We therefore conclude that there is no obvious discrepancy with the simulation produced for the purposes of comparison produced here and the results of TAH15 and PAO17. In future work, it will be necessary to compare the neutrino fluxes resulting from the use of different models to see whether the predictions are stable.

\begin{table}[!htbp]
   \caption{Comparison to Previous Results For Source Index $n=0$. See text for details.}
   \centering
   \begin{tabular}{@{} cccccccc @{}} 
      \toprule
               & $\alpha$ & $\log_{10}\left(\frac{E_\mathrm{max}}{\mathrm{eV}}\right)$   &$f_\mathrm{p}$(\%) & $f_\mathrm{He}$(\%) & $f_\mathrm{N}$(\%) & $f_\mathrm{Si}$(\%) & $f_\mathrm{Fe}$(\%)\\
                         \midrule
        RA18   & $[-0.22,+0.24]$ & $[18.69,18.77]$ & $[7,60]$ & $[5,74]$ &  $[25,59]$ & $[2,6]$ & $[0,0.2]$ \\
        TAH15  & $[+0.92,+1.74]$ & $[18.67,19.22]$ & $[1,33]$ & $[3,37]$ &  $[29,66]$ & $[0,33]$ & $[1,9]$ \\
        \underline{PAO17}  &     &                 &          &          &                       & \\
        SPG &    $[+0.81,+1.05]$ & $[18.62,18.70]$ & $[0,22]$ & $[45,71]$ & $[16,36]$ & $[4,6]$ & -- \\
        w/ sys & $[+1.20,+1.38]$ & $[18.49,18.59]$ & $[0,19]$ & $[19,48]$ & $[31,52]$ & $[5,15]$ & -- \\
        CTG$^1$    & $[-1.33,-0.68]$   & $[18.17,18.26]$ & $68$ & $31$& $1$ & $0.06$ & --\\
        CTG$^2$    & $[+0.81,+0.93]$   & $[18.60,18.64]$ & $0$ &  $0$& $88$ & $12$ & --\\
      \bottomrule
   \end{tabular}
   \label{tab:comparison}
\end{table}

\bibliographystyle{elsarticle-num}
\bibliography{<your-bib-database>}

\end{document}